\documentclass[11pt]{article}

\usepackage{amssymb,latexsym,amsmath}
\usepackage{xcolor}
\usepackage{bm}

\usepackage{hyperref}

\begin{document}

\newtheorem{theorem}{Theorem}[section]

\newtheorem{proposition}[theorem]{Proposition}

\newtheorem{lemma}[theorem]{Lemma}

\newtheorem{corollary}[theorem]{Corollary}

\newtheorem{definition}[theorem]{Definition}

\newtheorem{remark}[theorem]{Remark}

\newtheorem{exempl}{Example}[section]

\newenvironment{exemplu}{\begin{exempl}  \em}{\hfill $\surd$

\end{exempl}}

\newcommand{\ea}{\mbox{{\bf a}}}
\newcommand{\eu}{\mbox{{\bf u}}}
\newcommand{\ep}{\mbox{{\bf p}}}
\newcommand{\ed}{\mbox{{\bf d}}}
\newcommand{\eD}{\mbox{{\bf D}}}
\newcommand{\eK}{\mathbb{K}}
\newcommand{\eL}{\mathbb{L}}
\newcommand{\eB}{\mathbb{B}}
\newcommand{\ueu}{\underline{\eu}}
\newcommand{\ueo}{\overline{u}}
\newcommand{\oeu}{\overline{\eu}}
\newcommand{\ew}{\mbox{{\bf w}}}
\newcommand{\ef}{\mbox{{\bf f}}}
\newcommand{\eF}{\mbox{{\bf F}}}
\newcommand{\eC}{\mbox{{\bf C}}}
\newcommand{\en}{\mbox{{\bf n}}}
\newcommand{\eT}{\mbox{{\bf T}}}
\newcommand{\eV}{\mbox{{\bf V}}}
\newcommand{\eU}{\mbox{{\bf U}}}
\newcommand{\ev}{\mbox{{\bf v}}}
\newcommand{\eve}{\mbox{{\bf e}}}
\newcommand{\uev}{\underline{\ev}}
\newcommand{\eY}{\mbox{{\bf Y}}}
\newcommand{\eP}{\mbox{{\bf P}}}
\newcommand{\eS}{\mbox{{\bf S}}}
\newcommand{\eJ}{\mbox{{\bf J}}}
\newcommand{\leb}{{\cal L}^{n}}
\newcommand{\eI}{{\cal I}}
\newcommand{\eE}{{\cal E}}
\newcommand{\hen}{{\cal H}^{n-1}}
\newcommand{\eBV}{\mbox{{\bf BV}}}
\newcommand{\eA}{\mbox{{\bf A}}}
\newcommand{\eSBV}{\mbox{{\bf SBV}}}
\newcommand{\eBD}{\mbox{{\bf BD}}}
\newcommand{\eSBD}{\mbox{{\bf SBD}}}
\newcommand{\ecs}{\mbox{{\bf X}}}
\newcommand{\eg}{\mbox{{\bf g}}}
\newcommand{\paromega}{\partial \Omega}
\newcommand{\gau}{\Gamma_{u}}
\newcommand{\gaf}{\Gamma_{f}}
\newcommand{\sig}{{\bf \sigma}}
\newcommand{\gac}{\Gamma_{\mbox{{\bf c}}}}
\newcommand{\deu}{\dot{\eu}}
\newcommand{\dueu}{\underline{\deu}}
\newcommand{\dev}{\dot{\ev}}
\newcommand{\duev}{\underline{\dev}}
\newcommand{\weak}{\rightharpoonup}
\newcommand{\weakdown}{\rightharpoondown}
\renewcommand{\contentsname}{ }

\title{On the information content of the difference from hamiltonian evolution}
\author{\href{http://imar.ro/~mbuliga/}{Marius Buliga} \\ Institute of Mathematics of the Romanian Academy}
\date{Version: 12.02.2019. 
Licence: \href{https://creativecommons.org/licenses/by/4.0/}{CC BY 4.0} }


\maketitle

\begin{abstract}
A dissipative version of hamiltonian mechanics is proposed via a principle of minimal information content of the deviation from hamiltonian evolution. We show that we can cover viscosity, plasticity, damage and unilateral contact. 
\end{abstract}

\section{Introduction}

In hamiltonian mechanics a physical system is described by a state vector $q\in X$ and a momentum vector $p \in Y$. The evolution in time of the system is governed by a hamiltonian function $H=H(q,p,t)$, via the equations: 
\begin{equation} \left\{   
\begin{array}{rcl}
\dot{q} & = & \frac{\partial H}{\partial p}  (q,p,t) \\
- \dot{p} & = &  \frac{\partial H}{\partial q}  (q,p,t)
\end{array}
\right.
\label{hamilton}
\end{equation} 
where $\displaystyle \dot{q}$, $\displaystyle \dot{p}$ denote derivatives with respect to time. The evolution  is reversible.

We propose a dissipative modification of these equations, which extends the work from \cite{bham}, \cite{MBGDS1}, 
\cite{sben1}.

\begin{definition}
Given a hamiltonian $H$ and a likelihood function $\pi$ 
$$\displaystyle H: X \times Y \times \mathbb{R} \rightarrow \mathbb{R} \quad , \quad \pi: (X \times Y)^{3} \rightarrow [0,1]$$ 
a physical system described by the state vector $q \in X$ and the momentum vector $p \in Y$ evolves such that the  vector $\displaystyle \eta = (\eta_{q}, \eta_{p}) \in X \times Y$ defined by: 
\begin{equation} \left\{   
\begin{array}{rcl}
\dot{q} & = & \frac{\partial H}{\partial p}  (q,p,t) \, + \, \eta_{q} \\
- \dot{p} & = &  \frac{\partial H}{\partial q}  (q,p,t) \, + \, \eta_{p} 
\end{array}
\right.
\label{hamiltongap}
\end{equation} 
called the difference (or gap) from hamiltonian evolution driven by  $H$, maximizes the likelihood: 
\begin{equation}
\pi(z, \dot{z}, \eta) \, = \, \max_{z" \in X \times Y} \pi(z, \dot{z}, z")
\label{likely}
\end{equation} 
with the notations $\displaystyle z=(q,p) \in X \times Y$, $\displaystyle \dot{z} = (\dot{p}, \dot{q}) \in X \times Y$. 
\label{mainproblem}
\end{definition}

The likelihood function $\pi$ cannot be taken arbitrarily, otherwise the proposed modification would be too vague. We propose the following constraints on the likelihood function. 

First we define the information content function associated to the likelihood $\pi$ as  
\begin{equation}
I: (X \times Y)^{3} \rightarrow [0,+\infty] \, \, , \, \, I(z,z',z") \, = \, - \ln \pi(z, z', z")
\label{icfun}
\end{equation}
with the convention that $\displaystyle - \ln 0 \, = \, + \infty$.

\begin{definition}
The likelihood $\pi$ and the associated information content $I$ satisfy: 
\begin{enumerate}
\item[(a)] The information content function (\ref{icfun}) is convex in each of the 2nd and 3rd variables and it has the needed degree of smoothness required (for example it is lower semi-continuous with respect to the relevant topologies on $X$ and $Y$)  
\item[(b)] for any $\displaystyle z, z' \in X \times Y$,  the following maxima exist 
$$\displaystyle \max_{z" \in X \times Y} \pi(z, z', z")  \, , \, \max_{z" \in X \times Y} \pi(z, z", z')$$ and they are either $0$ or $1$.
\end{enumerate}
\label{likelyprop}
\end{definition}

The equation (\ref{likely}) can be rephrased as: given the hamiltonian $H$ and the information content function $I$, the physical system evolves such that at any moment it minimizes the information content of the gap from a hamiltonian evolution. Indeed, an evolution of the system is a curve $\displaystyle c_{0} :[0,T] \rightarrow X \times Y$ with the property that it minimizes the information content gap functional: 
\begin{equation}
G(c) =  \int_{0}^{T} I\left( c(t),\dot{c}(t),\dot{q}(t) - \frac{\partial H}{\partial p} (c(t),t), - \dot{p}(t) - \frac{\partial H}{\partial q}  (c(t),t) \right) \mbox{d}t
\label{icgap}
\end{equation}
among all admissible evolution curves $c(t) = (q(t), p(t))$.

\vspace{.5cm}

\paragraph{Pure Hamiltonian evolution.} Let's pick the information content (\ref{icfun}) to be: 
$$I(z, z', z") \, = \, \chi_{0} (z") = \left\{ \begin{array}{ll}
0 & \mbox{ if } z" = 0 \\ 
+\infty & \mbox{ otherwise } 
\end{array} \right.
$$
This corresponds to a likelihood function: 
$$ \pi(z,z',z") \, = \, \left\{ \begin{array}{ll}
1 & \mbox{ if } z" = 0 \\ 
0 & \mbox{ otherwise } 
\end{array} \right.
$$
The maximization of the likelihood (\ref{likely}) implies that the gap vector $\displaystyle \eta = 0$, therefore the evolution equations (\ref{hamiltongap}) reduce to the pure Hamiltonian evolution equations (\ref{hamilton}). 

\vspace{.5cm}

This example is trivial, we need a method to construct more interesting ones. One such method is based on the following observation, adapted from \cite{gds4}, section 2. We use the notations explained in section \ref{notations}, in particular we use the duality $\displaystyle \langle \langle \cdot , \cdot \rangle \rangle$ and we suppose that we have on $X, Y$ a topology compatible with it, so that the information content function $I$ is lower semicontinuous (lsc).

\begin{proposition}
The information content function (\ref{icfun}) satisfies the conditions  from Definition \ref{likelyprop}  if and only if the function  
\begin{equation}
b: (X \times Y)^{3} \rightarrow \mathbb{R} \cup \left\{ +\infty \right\} \, \, , \, \, b(z,z',z") \, = \ I(z,z', z") \, + \, \langle \langle z', z" \rangle \rangle 
\label{bipoic}
\end{equation} 
is a bipotential, i.e. it satisfies: for any $z \in X \times Y$,  
\begin{enumerate}
\item[(a)] for any $z', z" \in X \times Y$ the functions $b(z, z', \cdot)$ and $b(z, \cdot, z")$  are  convex (and lsc), 
\item[(b)] for any $z', z" \in X \times Y$ we have the equivalences 
\begin{equation}
z' \in \, \partial b(z, z', \cdot) (z") \, \Longleftrightarrow \, z" \in \, \partial b(z, \cdot, z") (z') \, \Longleftrightarrow \, I(z, z', z") = 0
\label{bipoequiv}
\end{equation}
where "$\partial$" denotes a subgradient, see section \ref{notations} for notations. 
\end{enumerate}
\label{biposync}
\end{proposition}

Bipotentials were introduced in \cite{gds2} as a convex analysis notion which is well adapted for applications to non-associated constitutive laws. Bipotentials were used in soil mechanics,  plasticity, damage or friction. For the theory of bipotentials see the review paper \cite{gds3}. 

By concentrating our attention to the function $b$, instead of the information content $I$, we can build a host of examples. Indeed, for any lsc and convex function
$$\Phi: X \times Y \rightarrow \mathbb{R} \cup \left\{ +\infty \right\}$$
the associated function 
\begin{equation}
b(z, z', z") \, = \, \Phi(z') + \Phi^{*}(z")
\label{sepabipo}
\end{equation}
satisfies the conditions  (a), (b) from Proposition \ref{biposync} , where $\displaystyle \Phi^{*}$ is the polar, or Fenchel conjugate of $\Phi$. Indeed, the mentioned condition (b) is just a reformulation of the Fenchel inequality \cite{gds2}. 

\begin{corollary}
For an information content function of the form
\begin{equation}
I(z, z', z") \, = \, \Phi(z') + \Phi^{*}(z") \, - \, \langle \langle z', z" \rangle \rangle 
\label{sepaic}
\end{equation}
the equation(\ref{likely}) is equivalent with the symplectic Brezis-Ekeland-Nayroles principle  \cite{MBGDS1} definition 4.1,   \cite{sben1} definition 1.1: 
\begin{equation}
\eta \, \in \, \partial \Phi \left(\dot{z}\right)
\label{sbennew}
\end{equation}
\label{corsben}
\end{corollary}

\section{Examples}
\label{secex}

\paragraph{Viscosity, Rayleigh dissipation.} In particular, let's pick 
$$\Phi(z') \, = \, \Phi(q',p') \, = \, \phi(q')$$
where $\displaystyle \phi: X \rightarrow \mathbb{R} \cup \left\{ + \infty \right\}$ is a convex, lsc function. A straightforward computation of $\displaystyle \Phi^{*}$ gives: 
$$\Phi^{*}(z") \, = \, \Phi^{*}(q", p") \, = \, \sup \left\{ \langle \langle (q', p') , (q",p") \rangle \rangle \, - \, \phi(q') \, \mid \, q' \in X, p' \in Y \right\} \, = \, $$
$$   \, = \, \sup \left\{ \langle q', p" \rangle + \langle q", p' \rangle \, - \, \phi(q') \, \mid \, q' \in X, p' \in Y \right\} \, = \, $$
$$ \, = \, \chi_{0}(q") \, + \phi^{*}(p")$$
therefore the information content has the expression: 
$$ I(z,z', z") \, = \, \phi(q') \, + \, \phi^{*}(p") \, + \, \chi_{0}(q")\, - \langle q', p" \rangle $$
By the corollary \ref{corsben} we obtain the equations: 
\begin{equation} \left\{   
\begin{array}{rcl}
\dot{q} & = & \frac{\partial H}{\partial p}  (q,p,t)  \\
- \dot{p} & = &  \frac{\partial H}{\partial q}  (q,p,t) \, + \, \eta_{p} \\
\eta_{p} & \in & \partial \phi \left( \dot{q} \right)
\end{array}
\right.
\label{hamiltongapvis}
\end{equation} 
This shows that $\phi$ is a Rayleigh dissipation potential.

\vspace{.5cm}

\paragraph{Elasto-plasticity.} Take a hamiltonian system with space state $X$ and momentum space $Y$ and supplement the state and the momentum with a new pair of state and momentum spaces: 
$$(q, q_{I}) \, \in \, X \times X_{I} \, , \, (p, p_{I}) \in Y \times Y_{I}$$
Suppose further that the pair of spaces $\displaystyle (X_{I}, Y_{I})$ are in duality, so that we can define 
$$\langle (q, q_{I}), (p, p_{I}) \rangle \, = \, \langle q, p \rangle \, + \, \langle q_{I}, p_{I} \rangle$$
which leads us to a duality product of $\displaystyle \left( X \times X_{I} \right) \times \left( Y \times Y_{I} \right)$ with itself: 
$$ \langle \langle (q', q_{I}', p', p_{I}'), (q", q_{I}", p", p_{I}") \rangle \rangle \, = \, \langle q', p" \rangle + \langle q", p' \rangle + \langle q_{I}', p_{I}" \rangle + \langle q_{I}", p_{I}' \rangle$$
In this setting, we take $\displaystyle N = \left( X \times X_{I} \right) \times \left( Y \times Y_{I} \right)$ and 
$$ \Phi \left( q', q_{I}', p', p_{I}'\right) \, = \, \phi \left(p_{I}' \right)
$$
where $\displaystyle \phi: Y_{I} \rightarrow \mathbb{R} \cup \left\{ + \infty \right\}$ is a convex lsc function. 

There are two differences with respect to the viscosity example: there is a cartesian decomposition of the state space and momentum space, and the dissipation potential depends on (a component of) the momentum variable, while previously the dependence was on the state variable. 

By a computation analoguous with the one from the previous example we obtain: 
$$ \Phi^{*} \left( q", q_{I}", p", p_{I}"\right) \, = \, \phi^{*}\left( q_{I}"\right) \, + \, \chi_{0}\left( p" \right) \, + \, \chi_{0} \left( p_{I}" \right) \, + \, 
\chi_{0} \left( q" \right) $$
Via the corollary \ref{corsben} we obtain the equations: 
\begin{equation} \left\{   
\begin{array}{rcl}
\dot{q} & = & \frac{\partial H}{\partial p}  (q,p,q_{I}, p_{I},t)  \\
\dot{q}_{I} & = & \frac{\partial H}{\partial p_{I}}  (q,p,q_{I}, p_{I}, t) \, + \, \eta_{q,I}  \\
- \dot{p} & = &  \frac{\partial H}{\partial q}  (q,p,q_{I}, p_{I}, t)  \\
- \dot{p}_{I} & = &  \frac{\partial H}{\partial q_{I}}  (q,p,q_{I}, p_{I}, t)  \\
\eta_{q,I} & \in & \partial \phi \left(  \dot{p}_{I} \right)
\end{array}
\right.
\label{hamiltongapplas}
\end{equation} 
In particular, let's take $\displaystyle X = X_{I}$, $Y = Y_{I}$, Hilbert spaces, and a hamiltonian of the form: 
$$H(q,p,q_{I}, p_{I},t) \, = \, K(p) \, + \, E\left( q - q_{I} \right) \, - \, \langle q, f(t) \rangle$$
where $K(p)$ is the kinetic energy and $\displaystyle E\left( q - q_{I} \right)$ is the elastic energy. We denote the elastic force by 
$$ \sigma \, = \,  \frac{\partial E}{\partial q}  (q - q_{I})$$ 
The 2nd and the 4th equations of the system (\ref{hamiltongapplas}) give: 
$$ \eta_{q,I} =  \dot{q}_{I} \, , \, \dot{p}_{I} = \sigma$$
The system  (\ref{hamiltongapplas}) reduces to the familiar equations: 
\begin{equation} \left\{   
\begin{array}{rcl}
\dot{q} & = & \frac{\partial K}{\partial p}  (p)  \\
\dot{p} & = &  f(t) \, - \, \sigma   \\
\dot{q}_{I} & \in & \partial \phi \left( \sigma \right)
\end{array}
\right.
\label{hamiltongapst}
\end{equation}

\vspace{.5cm}

In the next example we encounter a small extension of (\ref{sepaic}). 

\paragraph{Damage.} We consider here a standard model of damage. In \cite{bham} we introduced a formulation of hamiltonian evolution with dissipation which is the ancestor of the one proposed in this article. Then we  applied that formulation to a more involved damage model, where we used an Ambrosio-Tortorelli functional \cite{ambtor} in the hamiltonian of the model. 

As in the elasto-plasticity example, we supplement the spaces $X$, $Y$ with a new pair $\displaystyle (q_{I}, p_{I})$. Here $\displaystyle q_{I} = d \in \mathbb{R}$ is a damage variable and $\displaystyle p_{I} = r \in \mathbb{R}$ is a conjugated variable to the damage one. 

Thefore we  have a duality product over  $\displaystyle N = \left( X \times \mathbb{R} \right) \times \left( Y \times \mathbb{R} \right)$ : 
$$ \langle \langle (q', d', p', r'), (q", d", p", r") \rangle \rangle \, = \, \langle q', p" \rangle + \langle q", p' \rangle +  d'  r"  + d", r' \rangle$$
We choose a hamiltonian 
$$H(q,p,d, r ,t) \, = \, K(p) \, + \, k(r) \,  + \, (1-d) E\left( q \right) \, - \, \langle q, f(t) \rangle$$
where $K$ is the kinetic energy, $k$ is a term akin to the kinetic energy, but for the variable $r$ and $E$ is the elastic energy. 

The difference from the previous examples is that the damage variable and the sign of $\displaystyle \dot{d}$  are constrained (i.e. damage cannot decrease): 
$$ d \in [0,1] \quad , \quad \dot{d} \geq 0$$
We have to include these constraints in the expression of the information content: 
$$I(q,d,p,r, \dot{q}, \dot{d}, \dot{p}, \dot{r}, \eta_{q}, \eta{d}, \eta_{p}, \eta_{r}) \, = \, \chi_{[0,1]}(d) \, + 
$$
$$ + \, \, \chi_{[0,+\infty)}(\dot{d})  \, + \,  Y \dot{d} \, + \, \chi_{(-\infty,Y]}(\eta_{r}) + \, $$
$$ + \, \chi_{0}(\eta_{p}) \, + \, \chi_{0}(\eta_{q}) \, + \, \chi_{0}(\eta_{d})  \, - \, \langle \langle (\dot{q}, \dot{d}, \dot{p}, \dot{r}), (\eta_{q}, \eta_{d}, \eta_{p}, \eta_{r}) \rangle \rangle $$
The expression of the information content does not have the form (\ref{sepaic}), but instead it has the form 
$$ I(z, \dot{z}, \eta) \, = \, \mbox{ { \it constraints }} \, + \, \Phi(\dot{z}) + \Phi^{*}(\eta) \, - \, \langle\langle \dot{z}, \eta \rangle\rangle $$
with, in this case
$$\Phi(\dot{z}) \, = \, \phi(\dot{d}) \, = \, \chi_{[0,+\infty)}(\dot{d})  \, + \,  Y \dot{d}
$$
We obtain the equations: 
\begin{equation} \left\{   
\begin{array}{rcl}
\dot{q} & = & \frac{\partial K}{\partial p} (p)  \\
\dot{d} & = & \frac{\partial k}{\partial r} (r)  \\
- \dot{p} & = &  (1-d) \frac{\partial E}{\partial q}  (q) \, - \, f(t)  \\
- \dot{r} & = &  - E(q) \, + \, \eta_{r} 
\end{array}
\right.
\label{hamiltongapdam}
\end{equation} 
and the inequations: 
\begin{equation} \left\{   
\begin{array}{rcl}
d & \in & [0,1]  \\
\dot{d} & \geq  & 0  \\
\eta_{r} & \in & \partial \phi \left(  \dot{d} \right)
\end{array}
\right.
\label{hamiltongapdamineq}
\end{equation}

\vspace{.5cm}

In the next example we shall need a smoothness assumption. The alternative would be to renounce at the condition \ref{likelyprop} (a) and to keep only the condition (b). 

\paragraph{Unilateral contact.} We take $X$, $Y$ to be dual, finite dimensional Banach spaces and $M \subset X$ to be a closed, nonempty set of admissible states $q$. With the notations about Fr\'echet normals from \cite{kruger} 1.4, we shall need that $M$ is smooth enough, in the sense that the  tangent cone at any $q \in M$ is the polar of the normal cone at x, denoted by $N(q \mid M)$. 

There is no special form of the hamiltonian required. The information content has the form: 
$$I(z, \dot{z}, \eta) \, = \, \chi_{M}(q) \, + \, \chi_{N(q\mid M)} \left( \eta_{p} \right) \, + \, \left( \chi_{N(q\mid M)} \right)^{*} \left(\dot{q}\right) \, - \, \langle\langle \dot{z} , \eta \rangle\rangle$$
The smoothness assumption on $M$ is that 
$$\left( \chi_{N(q\mid M)} \right)^{*} \, = \, \chi_{T(q\mid M)}
$$
i.e. that the tangent cone is the polar of the normal cone. In general it is true that the normal cone is the polar of the tangent cone, but not the other way around. An alternative would be to not make any smoothness assumptions on $M$ and instead to take an information content of the form 
$$I(z, \dot{z}, \eta) \, = \, \chi_{M}(q) \, + \, \chi_{N(q\mid M)} \left( \eta_{p} \right) \, + \, \chi_{T(q\mid M)} \left(\dot{q}\right) \, - \, \langle\langle \dot{z} , \eta \rangle\rangle$$
Again, in general the tangent cone is not convex, therefore this form of information content contradicts the convexity condition \ref{likelyprop} (a). 

We obtain the equations: 
\begin{equation} \left\{   
\begin{array}{rcl}
\dot{q} & = & \frac{\partial H}{\partial p} (q,p,t)  \\
- \dot{p} & = &  \frac{\partial H}{\partial q}  (q,p,t) \, + \, \eta_{p}
\end{array}
\right.
\label{hamiltongapcont}
\end{equation} 
and the unilateral contact conditions:
\begin{equation} \left\{   
\begin{array}{rcl}
q & \in & M  \\
\dot{q} & \in  & T(q\mid M)  \\
\eta_{p} & \in & N(q\mid M) \\
\langle \dot{q} , \eta_{p}  \rangle & = & 0
\end{array}
\right.
\label{hamiltongapcontineq}
\end{equation}


\section{Notations and useful definitions}
\label{notations}

\paragraph{General notations.} 
The space $\displaystyle X$ of states $q \in X$ and the space $\displaystyle Y$ of momenta $\displaystyle p \in Y$ are  real topological vector spaces in duality:  
$$ (q,p) \in X \times Y \, \mapsto \langle q , p \rangle \in \mathbb{R}$$
We suppose the usual: the duality is bilinear, continuous and for any linear and continuous functions $\displaystyle L: X \rightarrow \mathbb{R}$, $\displaystyle G: Y \rightarrow \mathbb{R}$ there exist $q \in X$, $\displaystyle p \in Y$ such that $\displaystyle L(\cdot) = \langle \cdot, p \rangle$ and $\displaystyle G(\cdot) = \langle q, \cdot \rangle$. 

The space  $\displaystyle N = X \times Y$ is in duality $\displaystyle \langle \langle \cdot , \cdot \rangle \rangle : N \times N \rightarrow \mathbb{R}$ with itself by: 
$$ \langle \langle (q_{1}, p_{1}) , (q_{2}, p_{2}) \rangle\rangle \, = \, \langle q_{1}, p_{2} \rangle \, + \, \langle q_{2}, p_{1} \rangle
$$

The space $N$ is also symplectic, with the symplectic form defined by: for any $\displaystyle z_{1} = (q_{1}, p_{1})$ , $\displaystyle z_{2} = (q_{2}, p_{2})$ 
$$\omega(z_{1}, z_{2}) \, = \, \langle \langle J z_{1}, z_{2} \rangle \rangle \, = \, \langle q_{1}, p_{2} \rangle \, - \, \langle q_{2} , p_{1} \rangle $$

For any differentiable function $\displaystyle H: N \rightarrow \mathbb{R}$ the gradient of $H$ at a point $z \in N$ is the element $\displaystyle DH(z) \in N$  with the property that 
$$ \lim_{\varepsilon \rightarrow 0} \frac{1}{\varepsilon} \left( H \left( z + \varepsilon z'\right) - H(z) \right) \, = \, \langle \langle DH(z), z' \rangle \rangle
$$
and the symplectic gradient of $H$ is $XH(z) \in N$ is defined in a similar way by the equality 
$$ \lim_{\varepsilon \rightarrow 0} \frac{1}{\varepsilon} \left( H \left( z + \varepsilon z'\right) - H(z) \right) \, = \, \omega( XH(z), z' )
$$
If we use the partial derivatives notation 
$$ DH(q,p) \, = \, \left( \frac{\partial}{\partial p} H(q,p) , \frac{\partial}{\partial q} H(q,p)  \right)
$$
then 
$$ XH(q,p) \, = \, \left( \frac{\partial}{\partial p} H(q,p) , - \frac{\partial}{\partial q} H(q,p)  \right)
$$
With the introduction of the linear conjugation 
$$ \overline{\cdot} : N \rightarrow N \, \, , \, \, \overline{(q,p)} \, = , (q, -p)
$$
we have $\displaystyle XH(z) \, = \, \overline{DH(z)}$. 

\paragraph{Convex analysis notations.} These are the classical ones from Moreau \cite{moreau}.  We add $+ \infty$ the field of reals $\displaystyle \bar{\mathbb{R}} = \mathbb{R}\cup 
\left\{ +\infty \right\}$.  The addition operation is extended with  $a + (+ \infty) = + \infty$ for any 
$\displaystyle a \in \bar{\mathbb{R}}$. The multiplication with positive numbers is extended with: if $a>0$ then $a \left(+\infty \right) = + \infty$. 

For any function $\displaystyle \phi: X \rightarrow
\bar{\mathbb{R}}$, it's domain is $\displaystyle dom \, \phi = \left\{ x \in X \mbox{ : } \phi(x) \in
\mathbb{R} \right\}$. 

The set of lower semicontinuous (lsc), convex functions defined on $X$, with non-empty domain is $\displaystyle \Gamma_{0}(X)$. The indicator function $\displaystyle \chi_{A} \in \Gamma_{0}(X)$ of a convex and closed set  $A \subset X$ is  
$$\chi_{A} (x) = \left\{ \begin{array}{ll}
0 & \mbox{ if } x \in A \\ 
+\infty & \mbox{ otherwise } 
\end{array} \right. $$

For any natural number $n \geq 1$, a non empty set $A \subset X \times Y$ is $n$-monotone  monotone if for any collection $\displaystyle \left\{ (x_{k}, y_{k}) \in A 
\mbox{ : } k = 0, 1, ..., n\right\}$ we have the inequality: 
$$\langle x_{n} - x_{0} , y_{n}\rangle  + \sum_{1}^{n} \langle x_{k-1} - x_{k}, y_{k-1}
\rangle \, \geq 0 \quad .$$
The set $A \subset X \times Y$ is maximally $n$-monotone if it is $n$-monotone and maximal with respect to the inclusion of sets.  

The set $A$ is cyclically monotone if it is $n$ monotone for any natural number $n \geq 1$. It is cyclically maximal monotone if it cyclically monotone and maximal with respect to the inclusion of sets.

The subdifferential of a function $\displaystyle \phi: X \rightarrow
\bar{\mathbb{R}}$ at a point $x \in X$ is the set: 
$$\partial \phi(x) = \left\{ u \in Y \mid \forall z \in X  \  \langle z-x, u \rangle \leq \phi(z) - \phi(x) \right\} \  .$$ 

The polar of a function $\displaystyle \phi: X \rightarrow \bar{\mathbb{R}}$ is $\displaystyle \phi^{*}: Y \rightarrow \bar{\mathbb{R}}$
$$ \phi^{*} (y) \, = \, \sup \left\{ \langle x, y \rangle - \phi(x) \, \mid \, x \in X \right\} 
$$
The polar is always convex and lsc. 

Polars and subgradients are related by the Fenchel inequality. For any function $\displaystyle \phi: X \rightarrow \bar{\mathbb{R}}$ which is convex, lsc, we define  
$$ c(x,y) \, = \,  \phi(x) \, + \, \phi^{*}(y) \, - \, \langle x, y \rangle
$$
for any $x \in X$, $y \in Y$. The Fenchel inequality has two parts: 
\begin{enumerate}
\item[(a)] $\displaystyle c(x,y) \geq 0$
\item[(b)] $\displaystyle c(x,y) = 0 \, \Longleftrightarrow \, y \in \partial \phi(x) \, \Longleftrightarrow \, x \in \partial \phi^{*}(y)$
\end{enumerate}


\begin{thebibliography}{10}

\bibitem{ambtor} L. Ambrosio, V. Tortorelli, On the Approximation of Free Discontinuity Problems, Bollettino UMI 7, 6-B (1992), 105-123




\bibitem{bham} M. Buliga, Hamiltonian inclusions with convex dissipation with a view towards applications, {\it Mathematics and its Applications} {\bf 1}, 2 (2009), 228-251, \href{http://arxiv.org/abs/0810.1419}{arXiv:0810.1419}.





\bibitem{MBGDS1} M. Buliga, G. de Saxc\'e,  A symplectic Brezis-Ekeland-Nayroles principle, {\it Mathematics and Mechanics of Solids}  {\bf 22}, 6, (2017)

\bibitem{sben1} M. Buliga, A stochastic version and a Liouville theorem for hamiltonian inclusions with convex dissipation, \href{https://arxiv.org/abs/1807.10480}{arXiv:1807.10480}

\bibitem{gds3} M. Buliga, G. de Saxc\'e, C. Vall\' ee,  Bipotentials for non monotone multivalued operators: fundamental results and applications, Acta Applicandae Mathematicae, 110, 2(2010), 955-972 

\bibitem{gds4} M. Buliga, G. de Saxc\'e, C. Vall\' ee, Blurred maximal cyclically monotone sets and bipotentials  , Analysis and Applications 8 (2010), no. 4, 1-14



\bibitem{kruger} A. Ya. Kruger, On Fr\' echet subdifferentials, Journal of Mathematical Sciences 16 (2003), no. 3,  3325-3358




\bibitem{moreau} J.-J. Moreau, Fonctionnelles convexes (Rome: Istituto Poligrafico e zecca dello stato, 2003), p. 71, 97.


\bibitem{rocka1} R.T. Rockafellar, On the maximal monotonicity of 
subdifferential mappings, Pacific J. Math. 33, 1 (1970), 209-216





\bibitem{conf} A. Oueslati, A.D. Nguyen, G. de Saxc\' e (2017). A Symplectic Minimum Variational Principle for Dissipative Dynamical Systems. In: Nielsen F., Barbaresco F. (eds) Geometric Science of Information. GSI 2017. Lecture Notes in Computer Science, vol 10589. Springer, Cham


\bibitem{gds2} G. de Saxc\'e, Une g\' en\' eralisation de l’in\' egalit\' e de Fenchel et ses applications aux lois constitutives, C. R. Acad. Sci., Paris, S\' er. II 314 (1992) 125-129



\end{thebibliography}
\end{document}